\documentclass[jgr]{agutex}
\usepackage{amsmath}
\usepackage{graphicx,graphics}%,epsfig}
\authorrunninghead{DE BARROS ET AL.}
\titlerunninghead{LP event source mechanism at Mt Etna}
\usepackage{multirow}
\newcommand{\ve}[1]{\boldsymbol{#1}}

\newcommand{\degre}{$^\circ\,$}
\newcommand{\degrenospace}{$^\circ$}
\newcommand{\be}{\begin{equation}}
\newcommand{\ee}{\end{equation} }
\usepackage[usenames]{color}
\definecolor{gray}{rgb}{0.5,0.5,1}
\renewcommand{\emph}[1]{\color{gray}#1\color{black}\smallskip}
\authoraddr{C. J. Bean, School of Geological Sciences, University College Dublin, Belfield, Dublin 4, Ireland.}
\authoraddr{L. De Barros, School of Geological Sciences, University College Dublin, Belfield, Dublin 4, Ireland. (louis.debarros@ucd.ie)}
\authoraddr{I. Lokmer, School of Geological Sciences, University College Dublin, Belfield, Dublin 4, Ireland.}
\authoraddr{J.-P. M\'etaxian, LGIT, Universit\'e de Savoie-IRD-CNRS, 73376 Chamb\'ery, France. }
\authoraddr{G. S. O'Brien, School of Geological Sciences, University College Dublin, Belfield, Dublin 4, Ireland.}
\authoraddr{D. Patan\`e, INGV-Catania, Piazza Roma 2, 95125 Catania, Italia.}
\authoraddr{G. Saccorotti, INGV-Pisa, Via della Faggiola 32, 56126 Pisa, Italia.}
\authoraddr{L. Zuccarello, INGV-Catania, Piazza Roma 2, 95125 Catania, Italia.}

\begin{document}

\title{Source mechanism of Long Period events recorded by a high density seismic network during the 2008 eruption on Mt Etna}

\author{L. De Barros\altaffilmark{1}, I. Lokmer\altaffilmark{1}, C.J. Bean\altaffilmark{1}, G. S. O'Brien\altaffilmark{1},  G. Saccorotti\altaffilmark{2}, J.-P. M\'etaxian\altaffilmark{4}, L. Zuccarello\altaffilmark{1,3},   D. Patan\`e\altaffilmark{3}}

\altaffiltext{1}{School of Geological Sciences, University College Dublin, Dublin, Ireland.}
\altaffiltext{2}{INGV-Pisa, Pisa, Italia.}
\altaffiltext{3}{INGV-Catania, Catania, Italia.}
\altaffiltext{4}{LGIT, Universit\'e de Savoie-IRD-CNRS, Chamb\'ery, France.}
%\linenumbers*[1]
\begin{abstract}
129 Long Period (LP) events, divided into two families of similar events, were recorded by the 50 stations deployed on Mount Etna in the second half of June 2008. During this period lava was flowing from a lateral fracture after a summit strombolian eruption. In order to understand the mechanisms of these events, we perform moment tensor inversions. Inversions are initially kept unconstrained to estimate the most likely mechanism. Numerical tests show that unconstrained inversion leads to reliable moment tensor solutions because of the close proximity of numerous stations to the source positions. However, single forces cannot be accurately determined as they are very sensitive to uncertainties in the velocity model. Constrained inversions for a crack, a pipe or an explosion then allow us to accurately determine the structural orientations of the source mechanisms. Both numerical tests and LP event inversions emphasise the importance of using stations located as close as possible to the source.\\

Inversions for both families show mechanisms with a strong volumetric component. These events are most likely generated by cracks striking SW-NE for both families and dipping 70\degre SE (Family 1) and 50\degre NW (Family 2).  For Family 1 events, the crack geometry is nearly orthogonal to the dike-like structure along which events are located, while  for Family 2 the location gave two pipe-like bodies which belong to the same plane as the crack mechanism. The orientations of the cracks are consistent with local tectonics, which shows a SW-NE weakness direction. The LP events appear to be a response to the lava fountain occurring on the 10$^{th}$ of May, 2008 as opposed to the flank lava flow.
\end{abstract}

%\keywords{Etna, Long-Period events, location, plumbing systems}

\begin{article}
 \section{Introduction}
Mt Etna is an active 3330 m high stratovolcano located on the East coast of Sicily, Italy. An eruptive period began on the 10$^{th}$ of May 2008 with a powerful lava fountain in the South East Crater, one of the four main summit craters. An eruptive fissure opened on the 13$^{th}$ of May on the East flank of the volcano, in the ``Valle del Bove'' \citep[see e.g.,][]{Cannata2009}. The flank eruption stopped on July 6$^{th}$ 2009.\\

Long Period (LP) events recorded on Mount Etna have a period range between 0.75 and 5 s (i.e. 0.2-1.3 Hz). These events are commonly defined by a narrow frequency range and an emergent onset \citep[see e.g.,][for a general definition of LP events]{chouet03}. In the last few years LP events on Mt Etna have been analysed and located in many studies \citep{falsaperla02,saccorotti07,lokmer07b,lokmer07,patane08,Cannata09b}.  Sources of these events are usually found to be located a few hundred meters below the summit craters. LP events are repetitive, which suggests a repeating action of the same, non-destructive source process.\\

 Moment tensor inversion (MTI) has been performed on several volcanoes to quantify the source processes of these events. Most of these studies \citep[e.g.][]{Ohminato98,nakano03,chouet03a,kumagai02,lokmer07b,Jolly2010} suggest fluid-filled crack mechanisms, often accompanied by single forces. Such a mechanism is supported by forward modelling of fluid-filled resonator systems with various geometries, such as crack or pipe-like conduits \citep[e.g.][]{chouet85,chouet86,jousset04}. This resonance produces slow interface waves, also called crack waves, whose dispersive properties allow for the generation of low frequency events from relatively small sources \citep{chouet86,ferrazzini1987}. The trigger mechanism for this excitation still remains uncertain, but they are thought to be related to instabilities in the fluid motion \citep{Ohminato98,Rust2008,Neuberg2006,Gilbert2008}; The fluid can be magma \citep{Neuberg2006}, water or steam \citep{Cusano2008,kumagai05}, gas \citep{lokmer07b}, or mixtures of these fluids \citep{Ohminato98}. Considering the hypothesis of a mechanical interaction between the fluid and the solid phase \citep{ferrazzini1987,jousset03}, the strong link between LP activity and fluid dynamics implies that the characterisation of LP source mechanisms is fundamental to the understanding of  processes in magmatic systems.\\

To correctly reconstruct the moment tensor, the topography \citep{neuberg2000,jousset04,Obrien2009} and the velocity model  \citep{bean08,Cesca2008} play important roles. A poor knowledge of the velocity structure can lead to apparently stable but erroneous solutions, as it induces uncertainties in the Green's functions. This is particularly true for signals with frequencies above 0.2 Hz, whereas the larger wavelengths of the Very Long Period events \citep{chouet03a} make this approach more stable \citep{kumagai10}. This issue can largely be solved by using stations located very close to the source positions \citep{bean08,kumagai10}.\\
 
On Mt Etna, the first MTI of LP events was performed by \cite{lokmer07b} complemented by a full investigation of the LP properties \citep{saccorotti07,lokmer07} and MTI of synthetic data \citep{bean08}. These authors suggest that the source mechanism generating LP events consist of a subvertical crack striking NNW-SSE, with a gas ``pulsing'' excitation. However their dataset contains only one station located in the summit area; they suggested that a larger dataset recorded in the close proximity to the sources would help to better constrain the inversion. For this reason, a joint Irish (University College Dublin), French (Universit\'e de Savoie, Chamb\'ery) and Italian (Istituto Nazionale di Geofisica e Vulcanologia, Catania and Pisa) experiment was conducted on Mt Etna in early summer during the 2008 eruption. An exceptionally high density network of 50 broad-band stations (30 of them were located close to the summit), recorded LP events. \cite{debarros09} located the source positions of 129 selected events belonging to two different families sharing similar waveforms. They found shallow source locations in agreement with previous studies on Mt Etna \citep{lokmer07,Cannata09b}, but the high resolution locations allowed them to determine a temporal migration never observed before.\\

In this study we first present the dataset, and then use numerical tests to investigate the resolution and robustness of constrained \citep{nakano05,lokmer07b} and unconstrained inversion with the large number of stations available here. For real data, unconstrained and constrained inversions give similar solutions. The best solution is suggested to be a crack mechanism for both families. We then comment on the influence of the station distribution, noise, mislocation and mismodelling. The  interpretation of these cracks, striking SW-NE, are finally discussed in relation to the volcanic activity.

\section{Data} \label{secdata}
Mount Etna is one of the most active volcanoes in the world. Since 1995 a period of increased activity began, with eruptions occurring approximately every year. Eruptions are of two types: intermittent explosive eruptions  from the summit craters  and flank eruptions, with higher effusion rates, which originate from lateral fissures. Volcanic activity has been monitored since 2003 by a permanent network of broadband stations by the INGV Catania observatory.  In 2008, there was a powerful, explosive eruption on the 10$^{th}$ of May. An intense seismic swarm  occurred on the 13$^{th}$ of May, and the amplitude of tremors, which are long-duration signals related to the eruptive activity with frequency contents between 0.5 and 3 Hz on Mt Etna \citep{patane08}, strongly increased, just before the opening of an eruptive fracture \citep{Napoli2008,Cannata2009}. This fissure, striking approximately NW-SE, developed from the base of the North East Crater at about 3050 m. a.s.l. and rapidly extended downwards to 2500 m. a.s.l. into the sliding East flank of the volcano in the Valle del Bove. The eruption was marked at the beginning by strong Hawaiian activity, the effusion rate then decreased to reach a moderate rate that lasted until the 6$^{th}$ of July 2009.  At the same time degassing occurred, mainly in the NE crater. LP seismic activity was very high before and during the beginning of the eruption and sharply decreased in June and July 2008 \citep{patane08b}. The tremor location and the geochemistry show the arrival of primitive 	magma at the end of June 2008, without any change in the lava flow rate \citep{corsaro2009,patane08b}. \\

\begin{figure*}
%\hspace{-2cm}
\includegraphics[height=9cm]{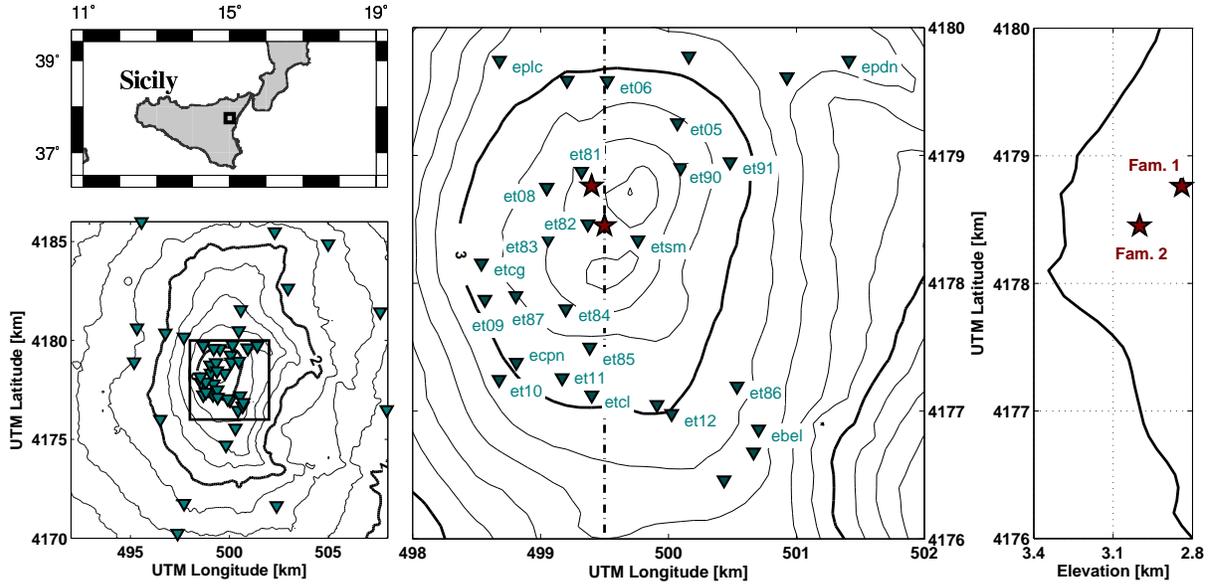}
\caption{Broadband station positions on Mt Etna. Left: Mt Etna location (top left panel) and all stations available on Mt Etna between the 18$^{th}$ of June and the 3$^{rd}$ of July 2008 (bottom left panel). Contour interval is 250 m. Middle: Summit area of the volcano with stations located within 2 km from the summit. Contour interval is 100 m. Stars indicate the average LP source location for both families found by \citet{debarros09}. Right: Same locations (stars) in a North-South cross-section at an UTM longitude of 499.5 km (marked by the dashed line in the middle panel).} \label{fig1}
\end{figure*}

A total of 50 stations (including the 16 permanent stations) with three-component broadband sensors (30, 40 or 60 s cut-off period), were
installed on Mt Etna between the 18$^{th}$ of June 2008 and the 3$^{rd}$ of July 2008. In particular, 30 of them were located at distances shorter than 2 km from the summit area (see Figure \ref{fig1}). \\
Before analyzing the data, we deconvolve the instrument response from the recorded signals. More than 500 events are found using a STA/LTA method on the bandpass (0.2-1.5 Hz) filtered data. We then classify these events using a cross-correlation analysis between all pairs of signals \citep{saccorotti07}. We keep the events that give a correlation coefficient greater than 0.9 with all other events on at least 3 out of the 4 permanent stations close to the summit.  We obtain 129 events which belong to two families \citep{debarros09}.\\

\begin{figure*}%[p]
%\hspace{-1cm}\vspace{-12cm}
\includegraphics[height=7cm]{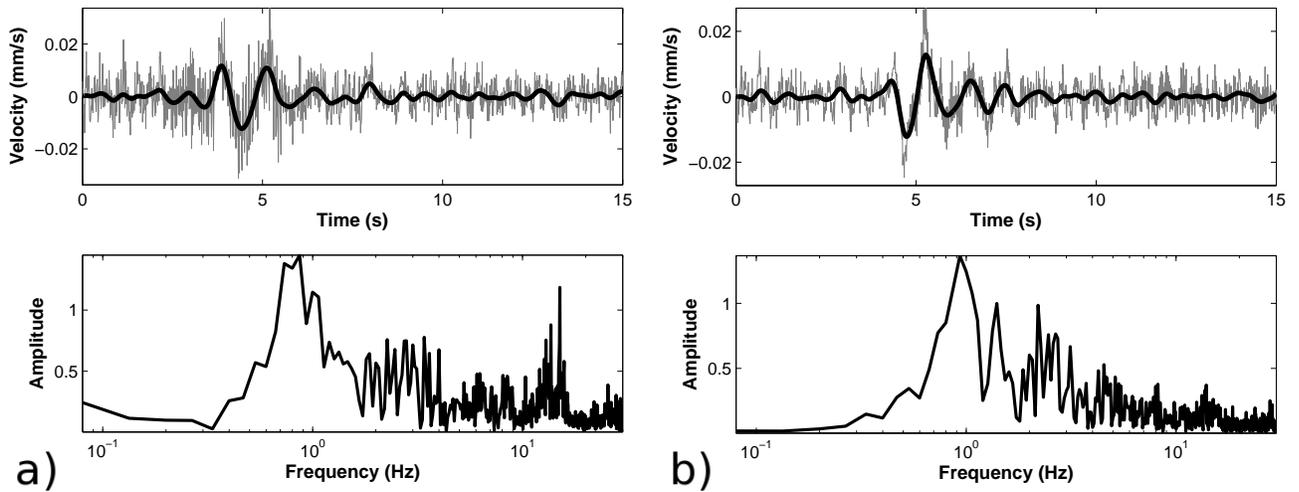} 
\caption{Data of an event from  a) Family 1 and b) Family 2, recorded at et81 station, vertical component. Top panels: waveforms (raw data and filtered data between 0.2 and 1.5 Hz) and lower panels: spectral content.}\label{fig2}
\end{figure*}

The first family (63 events) is only recorded in the first two days of the experiment (18$^{th}$-19$^{th}$ of June), while the second family (66 events) is distributed over the first four days. After the 22$^{th}$ of June, the amplitudes of the LP events decreases by an order of magnitude. In the same time period, the amplitudes of the tremors increase. Since both LPs and tremors are in the same spectral range, it is impossible to extract additional LP events due to poor signal-to-noise ratios. Most of the energy of the selected events is concentrated between 0.2 and 1.2 Hz, with a peak around 0.9 Hz. However, signal spectra show other peaks above 2 Hz, and some signals have higher frequency ($>$10 Hz) contents. For both families,  the waveforms (unfiltered and filtered between 0.2 and 1.2 Hz) and the spectral contents are shown in Figure \ref{fig2}. Although the waveforms are quite similar, the spectral peaks are not the same for the two families. Family 2 events exhibit a peak frequency slightly higher than in the case of Family 1. \\

\cite{debarros09} located the source of these events with a cross-correlation technique. Their analysis was based on the similarities between waveforms recorded by the different stations, which allowed them to measure the time delay between each pair of stations. They used a two-step technique by first determining an average location from the stack data of a given family and then relocating single events within that family. The robustness of this approach was checked on synthetic data. The source positions are located below the summit craters at very shallow depths, between 0 and 800m  from the summit for the first family and 0 and 400 m from the summit for the second family. The hypocenter positions are clustered into a subvertical, dike shaped structure striking NW-SE (Fam. 1) which branches into two pipe-like bodies (Fam. 2). The latter structure lies on a plane striking SW-NE and dipping 45\degre NW.  Some events from the two different families share the same location, thus the waveform difference between the two families has to be ascribed solely to a different source mechanism. However, the similarities of the waveforms indicate a repetitive source within each individual family.\\

\section{Method}
\subsection{Moment tensor inversion}
\subsubsection{Unconstrained MTI}
We performed a moment tensor inversion in the frequency domain as previously used by \cite{auger2006} and \cite{lokmer07b}. The $n^{th}$ component of the displacement field at station $s$ and at the frequency $\omega$, produced by a source located at the position $\ve{r}$ is denoted $u_n^s(\ve{r},\omega)$  and can be expressed as:
 \be \label{eq1}
  u_n^s(\ve{r},\omega)=G_{np,q}^s(\ve{r},\omega) ~ M_{pq}(\omega)+G_{np}^s(\ve{r},\omega) ~ F_{p}(\omega), \,\, \mbox{with} \,\, n,p,q=x,y,z,  
\ee 
$G_{np}^s(\ve{r},\omega)$ denotes the Green's functions (GF) and  $G_{np,q}^s(\ve{r},\omega)$ their spatial derivatives. We consider single forces ($F_{p}(\omega)$), and we assume a symmetric moment tensor (i.e. no rotational effects) with $M_{pq}(\omega)=M_{qp}(\omega)$. \\

Equation \ref{eq1} can then be rewritten in matrix form. The data $u_n^s(\ve{r},\omega)$  are merged in a column vector $\mathbf{d}$ and can be expressed in a linear form:
\be \label{eq1b}
\mathbf{d}=\mathbf{G}~ \mathbf{m}~,
\ee 
 where $\mathbf{G}$ is the matrix containing the Green's functions and their derivatives and $\mathbf{m}$ is a column vector of the moment tensor components and/or single forces. As we assume that the moment tensor is symmetric, only 6 moment tensor components and 3 single forces have to be determined for each frequency. The Source Time Function (STF) for each component is then obtained by applying an inverse Fourier transform.\\ 

The inversion problem is then linear, and equation \ref{eq1b} is solved for each frequency by a classical least-square  minimisation. The associated misfit of the waveforms is defined by
 \be\label{eq2}
\mathrm{Mis}=\frac{(\mathbf{d}-\mathbf{G}~ \mathbf{m})^T~~(\mathbf{d}-\mathbf{G}~ \mathbf{m})}{\mathbf{d}^T\mathbf{d}}.
\ee 
As we do not make any a priori assumptions about the mechanisms, the inversion is unconstrained and we can consider either the 6 moment components (MT) or the 6 moment components and 3 single forces (MT+F).  In these cases, we have 6 or 9 independent parameters to determine for each frequency in the frequency band of interest. \\

This inversion leads to 6 or 9 independent STFs, which need to be decomposed into mechanisms. The Principal Component Analysis \citep{vasco89}, which is based on a Singular Value Decomposition of the whole set of STFs, allows the estimation of a common STF and its contribution to each component, that is the scalar moment tensor.   However, this approach can be quite imprecise, especially if a time shift exists between the different components. The scalar moment tensor can also be determined using the peak-to-trough amplitude multiplied by the signum function \citep{chouet03a}. In both cases, the eigenvalues of the 3x3 scalar moment tensor give the mechanism and its eigenvectors yield the orientation of the principal axes. The solution can then be decomposed into the percentage of explosion, CLVD and Double Couple \citep{Vavrycuk01}. This last step, the decomposition, is however non-unique and unstable. 
\subsubsection{Constrained MTI}

\begin{figure*}
\includegraphics[height=6cm]{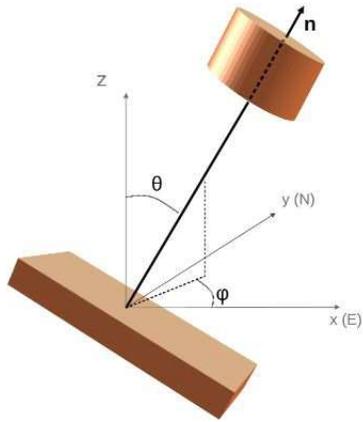}
\caption{Coordinate system used to define crack and pipe orientation.} \label{fig12}
\end{figure*}

\citet{nakano05} and \citet{lokmer07b} constrained the inversion to the particular mechanisms that are considered the most likely source mechanisms generating the LP events: a Crack (Cr), a Pipe (Pi) and an Explosion (Ex). A crack and a pipe are associated with a plane and a cylindrical shape, respectively, with the MT components ($M_{XX}$, $ M_{YY}$, $M_{ZZ}$)=(1,1,3) and (1,2,2) for a Poisson's ratio of 0.25, respectively. An explosion refers to a purely volumetric mechanism with eigenvalues (1,1,1). We use the sets of equations given by \citet{nakano05} to express the Cartesian components of the moment tensor as functions of the azimuth angle $\phi$ and dip angle $\theta$ of the symmetry axis (crack normal or longitudinal axis for a pipe, see Fig. \ref{fig12}). Moment tensor components are 1) for a crack:
\be \label{eqcr}
\left\{ \begin{array}{rcl} M_{XX}&=&M_0~ (\lambda/\mu + 2\sin^2\theta~ \cos^2\theta)\\
M_{YY}&=&M_0 ~(\lambda/\mu + 2\sin^2\theta~ \sin^2\theta)\\
M_{ZZ}&=&M_0 ~(\lambda/\mu + 2\cos^2\theta)\\
M_{XY}&=&M_0~ (\sin^2\theta~ \sin2\phi)\\
M_{XZ}&=&M_0~ (\sin2\theta ~\cos\phi)\\
M_{YZ}&=&M_0 ~(\sin2\theta ~\sin\phi),
\end{array} \right.
\ee 
2) for a pipe:
\be\label{eqPi}
\left\{ \begin{array}{rcl} M_{XX}&=&M_0~ (\lambda/\mu + \cos^2\theta~\cos^2\phi +\sin^2\phi)\\
M_{YY}&=&M_0~ (\lambda/\mu + \cos^2\theta~\sin^2\phi +\cos^2\phi)\\
M_{ZZ}&=&M_0~ (\lambda/\mu + \sin^2\theta )\\
M_{XY}&=&-1/2~ M_0 ~(\sin^2\theta ~\sin2\phi)\\
M_{XZ}&=&-1/2~M_0 ~(\sin2\theta ~\cos\phi)\\
M_{YZ}&=&-1/2~M_0~ (\sin2\theta ~\sin\phi),
\end{array} \right.
\ee 
and 3) for an explosion : 
\be \label{eqE}
\left\{ \begin{array}{rcl} M_{XX}&=&M_{YY}=M_{ZZ}=M_0 \\
M_{XY}&=&M_{XZ}=M_{YZ}=0
\end{array} \right.
\ee
$M_0$ denotes the seismic moment, $\lambda$ and $\mu$ are the Lam\'e's parameters. $X$, $Y$ and $Z$  refer to the East, North and vertical upward direction, respectively.  Herein, azimuth $\phi$ and dip $\theta$  are defined using the convention given in figure \ref{fig12}, i.e. $\phi$ is measured between 0 and 360\degre anticlockwise from East and $\theta$ is defined between 0 and 90\degre from the upward direction. It is worth stressing that the convention we use differs from the one used by \citet{nakano05}.\\

For each frequency, equation \ref{eq1b} can now be rewritten, as:
\be
\mathbf{d}=\mathbf{G_M}~\,~\mathbf{m_M}(\lambda/\mu,\theta,\phi)~\,~\mathrm{M_0}+\mathbf{G_F}~\,~\mathbf{m_F}~,
\ee
where $\mathbf{G_M}$ are the GF derivatives associated with the moment tensor components and $\mathbf{m_M}$ is the vector containing the moment tensor components as defined in equations \ref{eqcr}, \ref{eqPi} and \ref{eqE}, and $\mathrm{M_0}$ is the Source Time Function (STF) in the frequency  domain. $\mathbf{G_F}$ and $\mathbf{m_F}$ are the Green's functions and the source properties associated with the single forces, respectively. The last term of this equation refers to the inversion for single forces, and can either be included or omitted. If omitted, i.e. forces are not considered, inversion is denoted by Cr, Pi and Ex. As the vector $\mathbf{m_M}$ is independent of the frequency, for given values of $\lambda/\mu$, $\theta$, $\phi$ and $\omega$, the inversion procedure reduces to an inversion for a single parameter, $M_0(\omega)$. We search for the most likely solution by performing a grid search over the $\theta$-$\phi$ domain and making an inversion for the STF for each  $\theta$-$\phi$ pair. If single forces are considered (inversion denoted Cr+F, Pi+F and Ex+F), the moment tensor components are still determined using a grid search over the $\theta$-$\phi$ domain, while keeping the inversion for the forces unconstrained, as we do not assume any relationship between the single forces and the moment components. For given values of $\lambda/\mu$, $\theta$, $\phi$ and $\omega$, the number of unknowns to be determined is 3 for the single forces plus one for the moment components.\\

\subsection{Green's function (GF) computations}
\subsubsection{3D Elastic algorithm}
The Green's functions are computed using the 3D Elastic Lattice algorithm of \cite{obrien04}, based on a discrete particle scheme. The model is three-dimensional and includes topography of Mt Etna. It is centered on the volcano summit and has an area of 19.2x16x7 km with a 40 m grid size. Absorbing boundaries (4.8 km wide) are applied at the bottom and the edges of the model in order to prevent reflections from the model boundaries. The absorbing boundaries could be smaller in this study, but are designed to also work efficiently in the VLP frequency range, i.e. up to 0.02 Hz.  As the topography strongly distorts waveforms \citep{neuberg2000,jousset04,Obrien2009}, a free surface based on the Digital Elevation Model (DEM) of Mt Etna is required. The source function used for the GF computation is a Gaussian pulse with a 7.5 Hz cut-off frequency giving a flat spectrum below 1.5 Hz, as the signals are filtered in the 0.2-1.2 Hz range.  Below 1.5 Hz, the source signature is similar to the spectral signature of a Dirac delta function, which should theoretically be used to compute Green's functions. We have 25 grid points per minimum wavelength up to 1.5 Hz to prevent numerical dispersion for modelling including topography \citep{ohminato97}.\\ 

\subsubsection{Velocity model and source locations}
\cite{bean08} show that the moment tensor is very sensitive to incorrect velocity models, and particularly to shallow, low velocity structures; they also show that the effect of an erroneous velocity model  is stronger for stations further from the source. In this study, since 30 stations are located in the source near-field (less than one wavelength away) and we do not have any information on the shallow velocity properties, we choose to use a homogeneous model.  Velocities for $P-$ and $S-$waves are asummed to be 2000 m s$^{-1}$ and 1175 m s$^{-1}$, respectively. These velocities are similar to the results of the recent tomographic study of Mount Etna \citep{monteiller09}, which is homogeneous in the shallow part of the volcano, and to those determined in the location process of the LP events considered in this study \citep{debarros09}.  For stations close to the source, as shown by \citet{lokmer10}, the near-field effect has a strong influence on waveforms, but it is fully taken into account in our simulations.  Although attenuation is strong in volcanic structures \citep{jousset04}, it is not as important as scattering and topographic effects \citep{Obrien2009}. As the propagation distance is less than one wavelength and attenuation is also unknown in the shallowest part of the volcano, it is not considered here.\\

\cite{chouet03a} show the importance of a correct source location. However, as shown by \cite{lokmer07b}, in the presence of a poorly resolved shallow velocity model, there is a trade-off between  the source location  and source mechanism, i.e. coupled inversion can lead to an erroneous solution. Moreover, the GF calculation for multiple sources with such a large number of receivers is computationally expensive, for both direct and reciprocal approaches \citep{Eisner2001}, as GFs have to be computed from each source location or each receiver position. Hence we  use the source location from \cite{debarros09}.  The events do not share exactly the same source location, but we want to avoid the computation of the GFs for multiple sources. Moreover, the location can be slightly wrong because of the trade-off between the location and the velocity model.  We assume an average source position for both families described in section \ref{secdata} and we use these positions to compute the GFs required for the inversion of all LP events. In the following sections, we investigate the errors introduced by this assumption in synthetic tests and on real data.  The average source positions have UTM coordinates of (499.4, 4178.76, 2.84) for Family 1 and (499.5, 4178.45, 3.0) km for Family 2, i.e. 490 and 330 m below summit level (see Figure \ref{fig1}). \\

\section{Inversion of Synthetic Data}
\subsection{Numerical tests}
In the previous section, we introduced two strong assumptions which are i) a homogeneous model, and ii) an average source location for all events.  To assess the sensitivity of our inversion to uncertainties in the velocity structure, source mislocation and noise, we perform inversion of noisy numerical data computed with a velocity model and a source location different from those used in the Green's functions calculation. We also intend to assess: 1) if a constrained inversion gives more reliable results than an unconstrained inversion, and 2) if single forces have to be considered in the inversion. In the following sections, when dealing with numerical results, we use the notation $X$, $Y$ and $Z$ to refer to the East, North and vertical upward directions, respectively.\\

The velocity model used to compute the synthetic data is an artificial model consisting of a gradient with a $V_P$ increase from 1600 m s$^{-1}$ to 2500 m s$^{-1}$ from the surface to 500 m below the topography. The source location is misplaced by 90 m in the horizontal plane and 120 m vertically downward compared to the position where the GFs are computed. The source function is a Ricker wavelet, with 1 Hz central frequency. This analysis is carried out using synthetic data computed for two mechanisms: 1) Vertical crack, called $C_X$ with eigenvalues $M_{XX}=3~M_0$, $M_{YY}=M_0$, $M_{ZZ}=M_0$ (see eq.  \ref{eqcr}) and the crack-normal oriented along $X-$axis ($\phi$=0\degre and $\theta$=90\degre) and amplitude $M_0~=~3\,.~10^{12}$ N.m; 2) Same crack $C_X$ with an added single force $F_{45}$ with components $(F_X, F_Y, F_Z)~=~(9, 9, 9~\sqrt 2)~10^{9}$ N, which corresponds to an orientation of $\phi=45$\degre and $\theta$=45\degre. Random  noise is generated in the same spectral range as the waveforms. The amplitude of the noise added to the synthetics is chosen to be 25 \% of the maximum amplitude at etsm station, in order to achieve a noise level rougly similar to the one present in the real data.\\ 

\subsection{Results}
\begin{figure*}
\vspace{3.5cm}
\includegraphics[height=15cm]{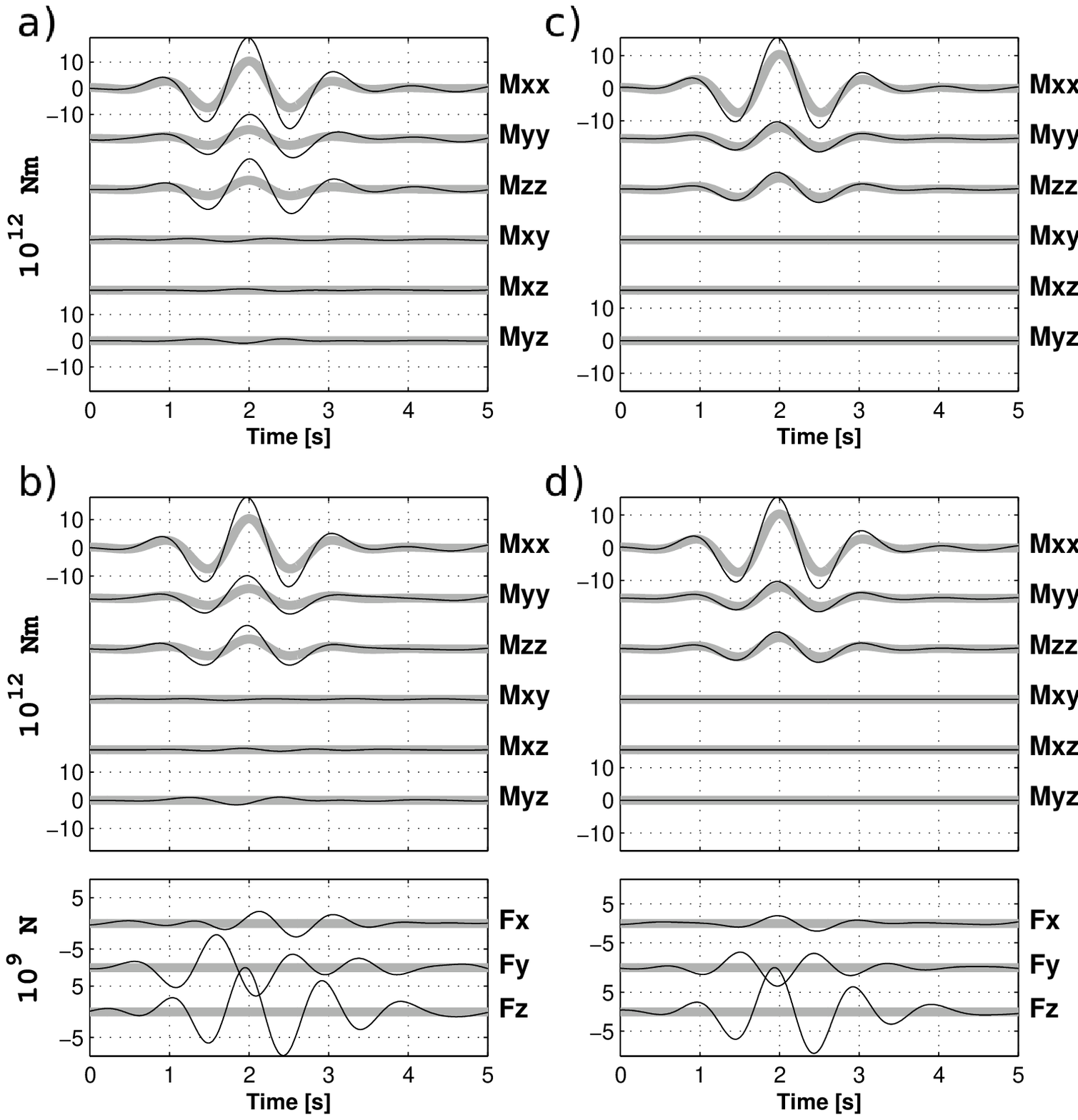}
\caption{Source Time Function (black thin lines) reconstructed using a) unconstrained  inversion for moment only (MT), b) unconstrained inversion for moment and single forces (MT+F), c) constrained inversion for crack only (Cr),  d) constrained inversion for crack and single forces (Cr+F). Noisy synthetic data are computed for the vertical crack $C_X$ without single forces, whose source mechanism is shown by the thick grey lines. In the figure, amplitude is $10^{12}$ N.m for the moment and $10^9$ N for the single forces.}\label{fig3}
\end{figure*}

For the data computed using  a crack $C_X$ without single forces, the mechanism and its orientation are reconstructed well by unconstrained inversion with, and without, single forces (see Figure \ref{fig3} and Table \ref{tab1}). The reconstructed moment however has a higher amplitude than the true solution. This is an expected result as i) the GF source is deeper (120 m) than the data source location and ii) the low velocity near the surface amplifies the signal relative to the GFs, calculated in a higher velocity. We then invert the synthetic data for a Crack, Pipe and Volumetric constrained mechanism (see Table \ref{tab1}). The minimum residual is obtained for the correct mechanism, i.e. for the crack constrained inversion. Its orientation and the STF are correctly recovered. As the solutions with and without noise are similar (not shown in figure), MTI does not appear to be very sensitive to the noise in this case. \\

The wrong velocity model and the mislocation produce spurious forces, which contribute to  about 50 \% in the reconstruction of the waveforms.  This can be due to i) the distortion of the waves propagating in the inhomogeneous velocity model and ii) difference in arrival times  for all stations when waveforms are computed with mislocation and wrong velocity models. We observe that moment tensors do not seem to be strongly affected by these factors for such a near-field deployment. The errors in the modelling seem to be accommodated by the single forces, i.e. the  inversion leads to spurious single forces to reconstruct the data when the modelling is not perfect.  Moreover, the spurious forces are very similar for both inversions presented in Figure \ref{fig3}, as they are representative of the model errors. In conclusion, i) constrained, or not, the inversion does not help to determine if single forces are real or due to mismodelling in the GF computations, and ii) unconstrained inversions are reliable for reconstructing the MT part of the source mechanisms for such a station network.\\

\begin{figure*}
\includegraphics[height=15cm]{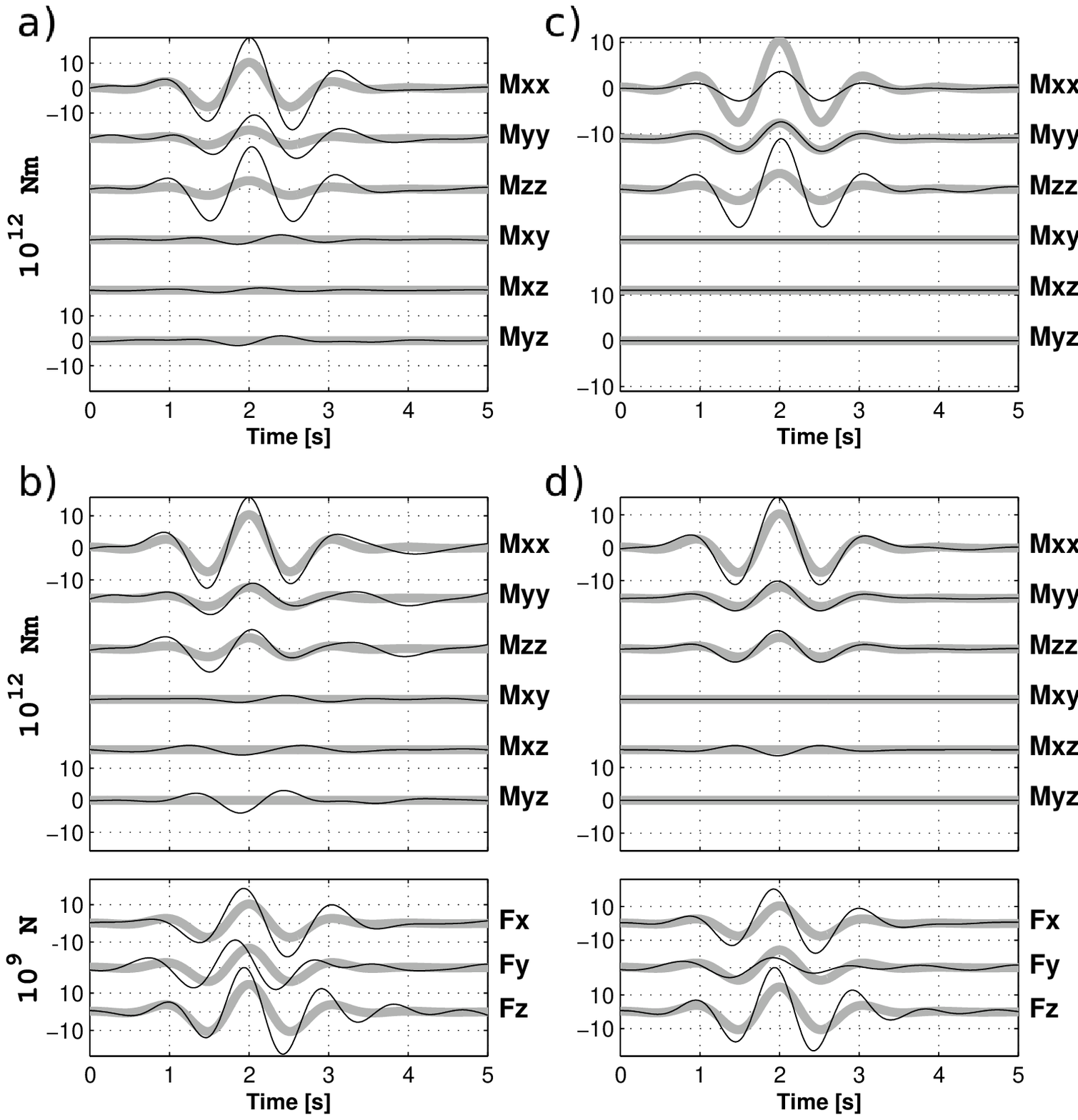}
\caption{Source Time Function (black thin lines) reconstructed using a) unconstrained  inversion for moment only (MT), b) unconstrained inversion for moment and single forces (MT+F), c) constrained inversion for crack only (Cr),  d) constrained inversion for crack and single forces (Cr+F). Noisy synthetic data are computed for the vertical crack $C_X$ and inclined single force $F_{45}$,  whose source mechanism is shown by the thick grey lines. In the figure,  amplitude is $10^{12}$ N.m for the moment and $10^9$ N for the single forces.}\label{fig4}
\end{figure*}

The same inversions are carried out for the second dataset where the true source is a crack  $C_X$ and a strong single force $F_{45}$ (see Fig. \ref{fig4} and table \ref{tab1}). The inversion for moment only (MT) gives an incorrect mechanism, similar to a pipe (1,1.8,2.2). The moment tensor components reconstructed by the unconstrained inversion (MT+F) are very close to the true solution. The orientation of the main axis is however close to the true solution for both cases (less than 15\degre error in dip and azimuth). For crack constrained inversions (Cr and Cr+F), moment is not reliable when forces are not considered as the reconstructed crack appears to be horizontal instead of vertical, while the inversion constrained for a crack and single forces (Cr+F) gives very good results.  Among the different geometries (Pi, Cr and E), the misfit minimum is not necessarily obtained for the correct mechanism (see table \ref{tab1}). As shown by \citet{lokmer10}, radiation patterns for Crack and Pipe mechanisms are very similar, and can appear the same if the wavefield sampling is not sufficiently dense.  When including single forces, the misfit minimum leads to a crack mechanism, but the misfit differences are still very small.  Similarly to the first case, the forces found in these inversion tests are not properly reconstructed as they include spurious forces due to velocity mismodelling and mislocation. The difference in misfit values between inversion with and without single forces is larger than for case 1, but it is not large enough to assure that forces are real. In conclusion, forces cannot be accurately reconstructed, but inversion with single forces is required if single forces exist.\\   

In order to make our results more general, we consider additional synthetic data computed for different crack geometry and single forces. The geometry of the crack called $C_L$ is inspired by the results of \citet{lokmer07b} and is defined by $\phi=35$\degre, $\theta=72$\degre and amplitude $M_0=3~\,~10^{12}$ N.m. We also consider a vertical force, designed by $F_Z$ with components ($0$,$0$,$6~\,~10^9$) N, as several authors \citep[e.g.][]{chouet03a} found a single force with this orientation. Data are then computed with the same conditions as in the previous tests, which include mislocation, mismodelling and noise. Results are summarized in table \ref{tab1}, for the different combination of the two cracks and the two forces. In agreement with the previous examples, the MTI is more accurate when single forces are included, as the solution obtained with the MT+F inversion is closer to the true mechanism than the solution reconstructed by the MT only inversion. The orientation of the major axis is however quite stable whatever the unconstrained inversion we used, the errors on the azimuths and dips are less than 20\degrenospace. Moreover, the Cr+F inversion leads to a very good solution in all cases, while the Cr solution is sometimes totally wrong. When single forces are not included, the misfit cannot be used to discriminate between the different geometries in constrained inversions, as the solution for pipe constrained inversion can have a smaller misfit than the one for a crack. \\

\subsection{Conclusion and strategy}
In the case of mismodelling and mislocation, moment tensor inversions do not allow single forces to be properly reconstructed. However, numerical tests show that the moment is more reliable if the inversion is carried out considering free single forces. Consequently, herein we allow single forces in the inversion to compensate for the errors coming from the velocity model and the source location. These single forces are however not considered for the interpretation of the mechanism.  A similar conclusion has been reached by \citet{Sileny2009}, who shows that, for  earthquakes with double-couple mechanisms with a small non-shear component, and in the presence of mislocation and mismodelling,  solutions are more stable when considering the 6 MT components than for a constrained double-couple inversion. Therefore, it can be sometimes better to have more unknowns in an inversion process in order to accommodate the errors. However, synthetic tests, like the ones presented here, are always necessary to  choose an inversion strategy (constrained MTI?, single forces?), as results will strongly depend on station density and topography. \\

With their stations configurations, \citet{bean08} showed that the strong sensitivity of the MTI to the shallow velocity model prevents the recovery of the correct unconstrained solution and can lead to totally wrong orientation of the mechanism. However, in the study presented here, as the mismodelling and mislocation effects seem to be balanced by using stations very close to the source, it is possible to correctly reconstruct the mechanism  using an unconstrained inversion. The interpretation of such a solution can be delicate, as the extraction of a single STF and the decomposition of the moment tensor components can be delicate and imprecise.  To help alleviate this problem, \cite{chouet2005} and \cite{kumagai05} first determine a rough approximation of the mechanism using an unconstrained inversion, and then refine the source characteristics by constraining the inversion. We choose to use a similar approach, which proceeds in two steps: 1) we invert for an unconstrained solution in order to determine the most reliable mechanism type (Crack, Pipe, Explosion), and 2) we use results of step 1 to perform a constrained inversion. The second step allows us to confirm  and check consistency of this solution and to refine the mechanism and its characteristics.  In both steps we consider forces and leave them unconstrained in order to accommodate the errors.\\

\section{Moment Tensor Inversion of Mt Etna Data}
We invert data from both families to determine the moment tensor using 16 stations (see section 6 for a justification of the number of stations) with the best azimuthal distribution and signal-to-noise ratio. Since some stations were not available at the beginning of the experiment, the set of stations is different for each family.  Individual events are contaminated by noise and do not share exactly the same source position \citep{debarros09}. The GFs are however computed for fixed positions (see section 3). To check this assumption (see discussion in section 6.2), we carried out inversion for 44 and 39 events for Family 1 and 2, respectively. The mean STF is obtained by averaging all the reconstructed STFs and the standard deviation gives us errors associated with noise and mislocation.  To do that, although the LPs have very similar amplitude, we normalise the STFs by the maximum of $M_{XX}$ in order to have comparable solutions for the different LPs. Errors for Family 1 (fig.  \ref{fig5}) are larger than for Family 2 (Fig. \ref{fig6}), but  the calculated STFs do not show any strong variations for either family.  As expected for LP multiplets, the source process is highly repetitive. Constrained inversion is then carried out using a single event in order to show that inversion of a single event is also stable.\\

\subsection{Family 1}
\begin{figure*}
\includegraphics[height=15cm]{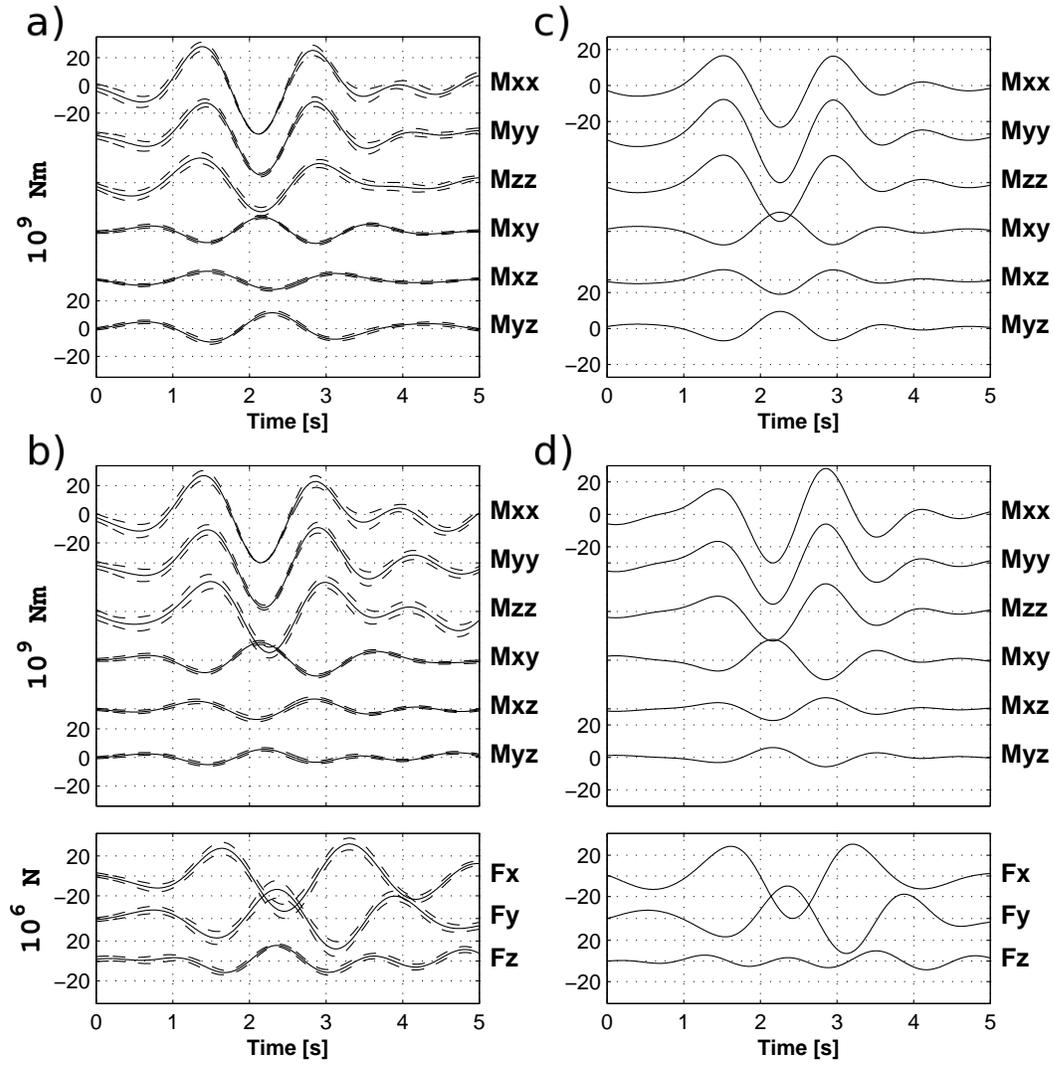}
\caption{MTI results for Family 1:  Mean solution and standard deviation for the Source Time Function obtained by the unconstrained inversion of 44 events for a) moment only (MT) and b) moment and single forces (MT+F); Source Time Function obtained by constrained inversion of a single event for c) crack constrained inversion (Cr); d) crack constrained inversion with single forces (Cr+F). Amplitude is $10^9$ N.m for the moment and $10^6$ N for the forces. }\label{fig5}
\end{figure*}

\begin{figure*}
\includegraphics[height=15cm]{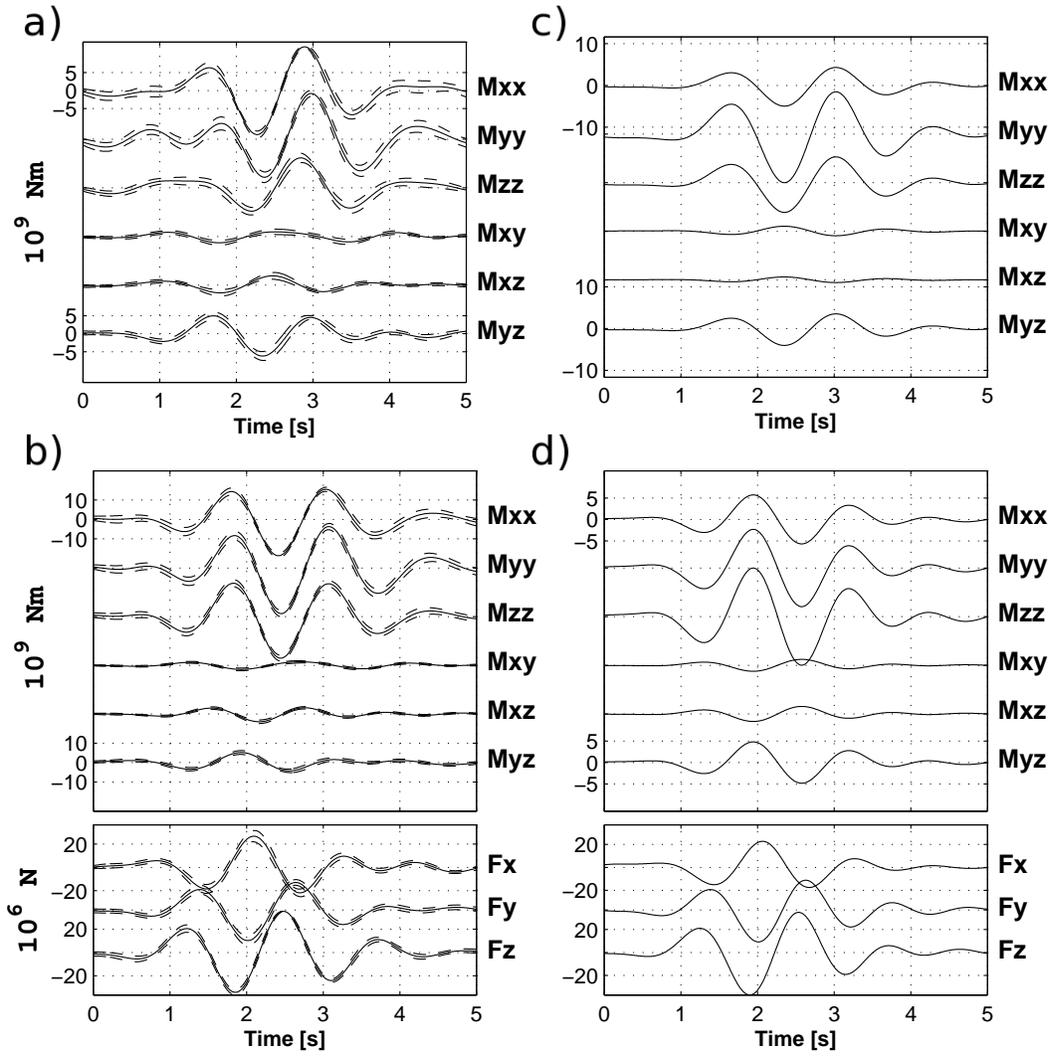}
\caption{MTI results for Family 2:  Mean solution and standard deviation for the Source Time Function obtained by the unconstrained inversion of 39 events for a) moment only (MT) and b) moment and single forces (MT+F); Constrained inversion of a single event for c) crack constrained inversion (Cr); d) crack constrained inversion with single forces (Cr+F). Amplitude is $10^9$ N.m for the moment and $10^6$ N for the forces.} \label{fig6}
\end{figure*}

\begin{figure*}
\includegraphics[height=10cm]{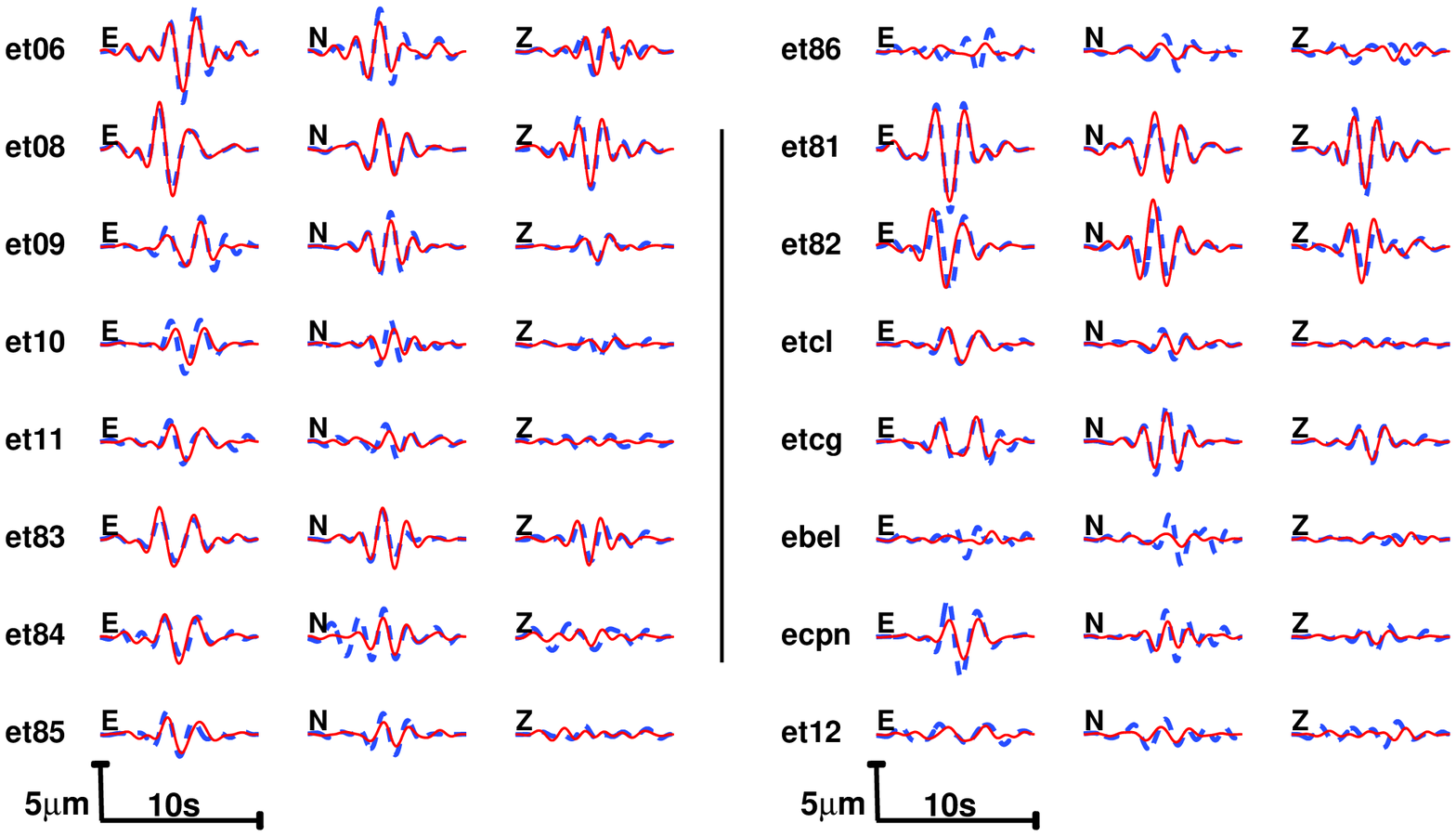}
\caption{Waveform (displacement) fit between the data (dashed lines) and the synthetic seismograms (continuous lines) for an individual event of Family 1. Inversion is unconstrained and includes single forces (MT+F). The misfit value is 27 \% (see Table \ref{tab2}).  }\label{fig7}
\end{figure*}

For Family 1, the STF reconstructed by unconstrained inversion (16 stations) with and without single forces are very similar (fig.  \ref{fig5}a and b). The misfit value (see table \ref{tab2}) is however considerably lower when forces are considered. Forces, whether physical or an artefact, do not change the moment tensor solution in this case. Waveform matches between data and reconstructed waveforms are shown in Figure \ref{fig7} for the 16 stations used for the unconstrained inversion with single forces for an individual event. Fits are very good for most of the stations very close to the source position. They disimprove for stations with lower amplitude signals due to the lower signal-to-noise ratio, and because the inversion gives more weight to the signals with the largest amplitude.  \\

\begin{figure*}
\includegraphics[height=8cm]{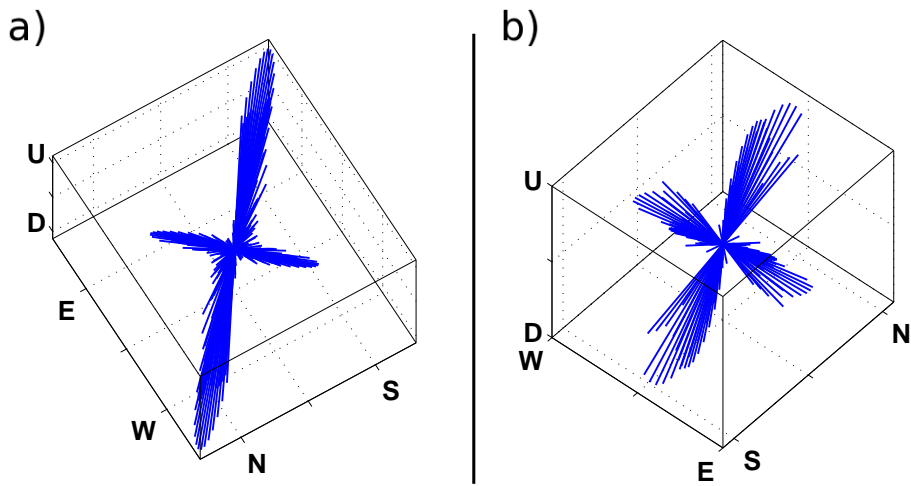}
\caption{Eigenvectors obtained for the MT+F inversion solutions for a) Family 1  and b) Family 2. Corresponding solutions are given in figure \ref{fig5}b for Fam. 1 and \ref{fig6}b for Fam. 2. Eigenvectors are sampled every 0.04 s when the amplitude of one of the components is greater than 50\% of the maximum amplitude.}\label{fig11}
\end{figure*}

The unconstrained inversions (MT and MT+F) lead to very similar mechanism and orientation. They can be interpreted as a crack (e.g., 1,1.1,2.3 for MT+F inversion). This is confirmed by the eigenvectors of the moment tensor solution, which are shown in figure \ref{fig11}.  In the second step, we invert for a crack solution, with and without single forces, and search for the azimuth and dip (Fig. \ref{fig5}c and d). Whereas STFs are slightly different for the four inversion results, they show a very similar amplitude and orientation of the crack mechanism. These results are in close agreement with the numerical tests shown in Figure 4. We can therefore assume that the single forces are probably not real or are too weak to be reconstructed. The moment tensor components show a crack whose normal is oriented with azimuth $\phi$=-40\degre and dip $\theta$=70\degre. \\

\subsection{Family 2}
\begin{figure*}
\includegraphics[height=10cm]{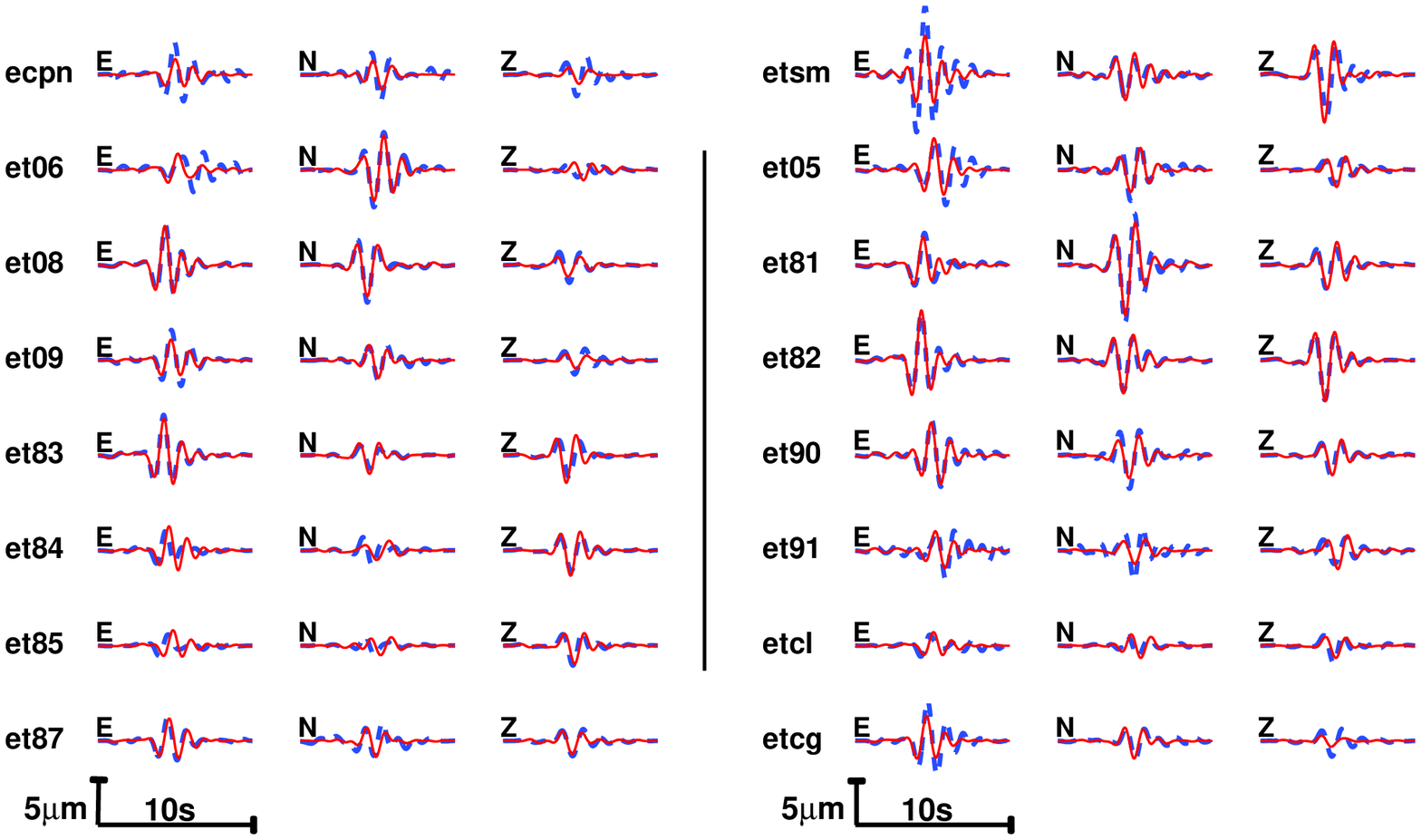}
\caption{Waveform (displacement) fit between the data (dashed lines) and the synthetic seismograms (continuous lines) for an individual event of Family 2. Inversion is unconstrained and includes single forces (MT+F). The misfit value is 21 \% (see tab. \ref{tab2}). }\label{fig8}
\end{figure*}

For Family 2, we also used 16 stations to perform MTI. The moment tensor solutions of the inversions with and without single forces are very different, both for the mechanism and for the STFs (see Figure \ref{fig6}). The misfit difference between these two inversions is very large (see Table \ref{tab2}). The waveforms cannot be properly explained without forces, though they are very well reconstructed when forces are considered (see Figure \ref{fig8}). By analogy with the numerical tests shown in Figure 5, we are more confident with the solution reconstructed with single forces.  However, the time shifts which exist among the  different moment components and single forces do not allow for an easy interpretation of the mechanisms. If we use a Principal Component Analysis (PCA) for the moment part \citep{vasco89}, the first principal component shows eigenvalues of (1,1.1,1.6), with nearly 80\% isotropic component.  The deviatoric part shows an axis pointing in the ($\phi$=110\degrenospace, $\theta$=50\degrenospace) direction. As the non-diagonal components of the tensor are shifted compared to the isotropic part and have a weaker amplitude, their effects are therefore underestimated by the PCA approach.  If we use only the maxima of the STFs, we find a source mechanism of (1,1.4,2.2) with the same orientation for the major axis. To have a clearer idea of the mechanism, we plot the eigenvectors of the MT+F solutions (Fig. \ref{fig11}). The eigenvectors are more spread out than for Family 1 but they however show a clear pattern for the mechanism, with a longer axis and two smaller axes with comparable length. This clearly indicates a mechanism similar to that of a crack. \\

To reconfirm the mechanism, we perform constrained inversions for a crack, a pipe and an explosion (see table \ref{tab2}). Smaller misfit values are obtained for the crack mechanism. Similarly to the unconstrained inversion, this solutions lead to a major axis in the $\phi$=110\degre and $\theta$=50\degre direction. The pipe and the explosive constrained inversions show  slightly higher misfits. We choose to discard these mechanisms as they do not show a major axis with an orientation consistent with the one obtained by the unconstrained inversion. This case is very similar to the inversion of synthetic data computed for a crack with a single force (Fig. 5). Part of these forces may be real but a significant portion is a consequence of model error. As shown by synthetic tests, it is impossible to accurately reconstruct the forces. For this reason, the single forces will not be quantitatively described and discussed.

\section{Discussion}
\subsection{Station distribution and density}
\begin{figure*}
\includegraphics[height=7cm]{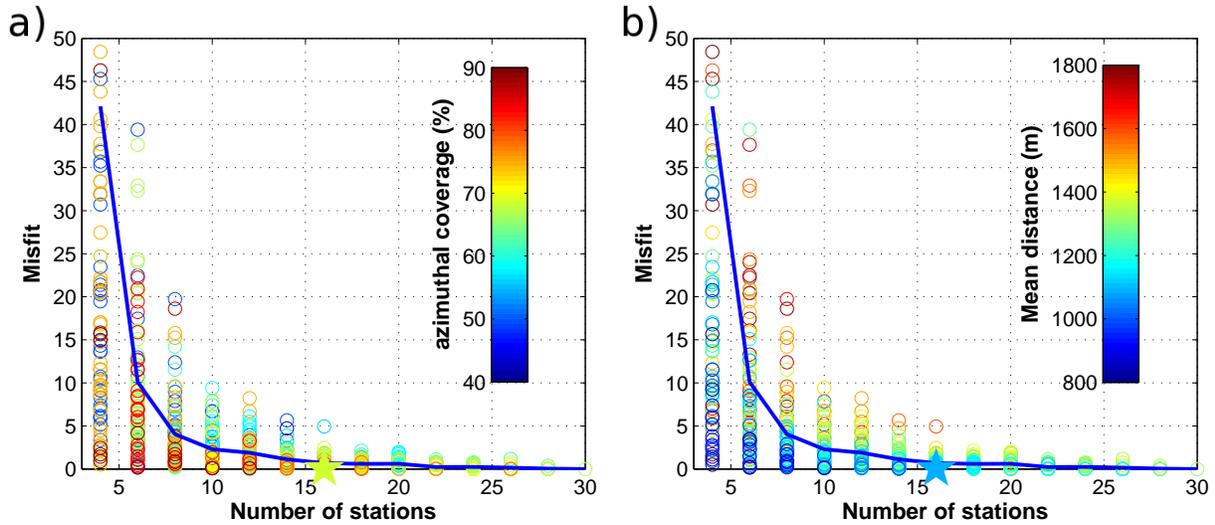}
\caption{RMS errors of the source time function versus the number of stations, for a single event from Family 2. The reference is the solution obtained when using the full set of 30 stations. The line show the mean. The stars refer to the inversion results shown in figure \ref{fig5}  and \ref{fig6}. Colorscale corresponds to a) azimuthal repartition in \% and b) average source-receiver distances in metre.}\label{fig10}
\end{figure*}

The major part of Family 1 events were recorded by only 16 stations. In order to have comparable results we chose to perform the inversion for both families using 16 stations. However,  Family 2 events and part of the Family 1 signals are recorded by 30 stations. We investigate the influence of the number of stations, azimuth repartition and propagation distance, by inverting a single event from Family 2  with different numbers of stations. For a number of stations $n$ between 4 and 28, we randomly choose up to 120 sets of stations, and perform an unconstrained MT inversion. We also compute the average propagation distance and the azimuthal coverage. The latter is estimated by dividing the azimuthal space into $n$ angular sectors with an angle of $360/n$\degre, and calculating the number of portions  in which at least one station is included. This value is then normalized by $n$ to have an estimation of the azimuth coverage quality as a percentage. The solution obtained using the 30 stations is taken as a reference, as this is the best solution we can expect from this dataset. We then look at the RMS error between the 6 moment components obtained using $n$ stations and the whole set of stations. Results are summarised in figure \ref{fig10}. We see that no matter how many stations are used in the inversion, it is still possible to have a solution close to the reference. However, the likelihoods vary significantly and at least 10 stations are required to be sure that the solution is not totally divergent from the reference. Propagation distance is clearly the crucial point, as the solutions diverge from the reference when the average distance increases. Sensitivity to the azimuth seems to be weaker;  the decrease of the misfit with azimuthal coverage  is less systematic. It is however never very bad, at least half of the azimuthal sectors are usually covered. Finally, the solution shown in figure \ref{fig5} and computed with 16 stations is very similar to the reference solution. \\

 In conclusion,  using between 10 and 30 stations we find that the number of stations does not change the solution very significantly as long as: 1) a broad azimuthal distribution is achieved and 2) some stations closest to the sources are used. This is mainly because the stations closest to the source strongly constrain the solution.  Using less than 8 stations, MTI usually leads to a different and certainly erroneous solution. This result, however, cannot be directly generalised to any other station distributions and any other volcano, but it should be tested using synthetic modelling for each individual case. Moreover, we do not use the same set of 16 stations for Family 1 and Family 2. We also verify that the difference between the solutions obtained for the two families is not produced by the station distribution difference. Inversion of Family 2 events leads to a similar result when using either set of stations.  \\ 

\cite{lokmer07b} show that the stations closest to the source help to constrain the source time function, while the others can be used to determine the mechanism. Here, the STF does not resemble the waveform recorded at the closest station (summit station etsm). Signals from this station show a complex waveform (see Fig. \ref{fig8}) probably due to local site effects and strong topographic effects. However, the other stations with small offset from the source display a signal very similar to the STF. In general, for stations close to the source, LP waveforms are not strongly distorted by propagation effects, which stabilises the STF reconstruction. In conclusion, MTI requires stations in close proximity to the source (near summit stations in our case) to be accurate, but not necessarily a dense network of stations (i.e. a minimum of 10 near summit stations in our case).

\subsection{Reliability of the solutions}
The moment tensor inversion solutions  for both families suggests a crack mechanism. Figure \ref{fig9} graphically summarizes the orientation of the solution for the moment components, with the orientation of the main location structures found by \cite{debarros09}. We are confident of the crack solution found for Family 1 as it appears stable for all of the inversion tests. For Family 2 the unconstrained solution shows a high volumetric component with a mechanism between a crack and an explosion. The constrained inversions and the eigenvector plot confirm the crack solution, therefore we have confidence in this solution. \\

The principal moment component for both families are about (1,1,2). To obtain these values for a crack, we need a Poisson's ratio of $\nu=1/3$, which implies $\lambda=2~\mu$. This high ratio is classically related to the high temperature in volcanic rocks \citep{chouet03a}. Fractured and unconsolidated media can also present a high Poisson's ratio \citep[e.g.][]{bourbie86}, which is most likely the case in the near subsurface of volcanoes. Another possible explanation of the difference with the theoretical crack mechanism (1,1,3) is that the latter is computed for an idealised point source with no realistic boundary conditions between the cracks perimeter and the surrounding medium. We also note the presence of strong single forces, especially for Family 2. If some of these forces are real they could be related to mass movement or drag forces \citep{Takei1994}. It also could mean that the mechanism is more complex than can be solved by the MTI used here as: 1) the source can comprise of several time-delayed mechanisms, 2) it can have a spatial extent with complex geometry, and 3) it involves rotational effects, such as torque. In these cases, mechanisms cannot be described by a second-order symmetric moment tensor and require a higher order moment tensor or a different approach to be accurately determined.  Numerical experiments have yet to be conducted  to investigate how to recover complex sources such as pressure dipoles, torque effects, etc. \\

Small time shifts between the different components of the solution and strong single forces can be partly explained by the source mislocation. For the events used in this study the common source locations  are associated with standard errors on the location of 150 m and 65 m for Fam. 1 and 2., respectively \citep{debarros09}. These errors are smaller than the one used for the synthetic tests (175m), which does not prevent the reconstruction of accurate MT solutions. Furthermore, the standard deviations of the solutions (Fig. \ref{fig5} and \ref{fig6}) are very small; this shows that the solution is not strongly affected by the choice of the common source locations.  Mismodelling of the velocity structure is another reason for the strong single forces and time shifts. Lateral velocity variations have not been considered, either in the synthetic test or in the GF modelling. They can however be strong, notably because of the deconsolidation of the East flank of the volcano, which is sliding. The velocity contrast is certainly stronger in the shallowest part of the volcano, which can lead to greater errors for the shallowest family (Fam. 2). Finally, a third explanation for the errors  in the solution can be the complexity of the source; however, being limited by the accuracy of the velocity model, it is impossible to unambiguously solve the issue of the source complexity itself.\\

\subsection{Source mechanisms}
\begin{figure*}
\includegraphics[height=12cm]{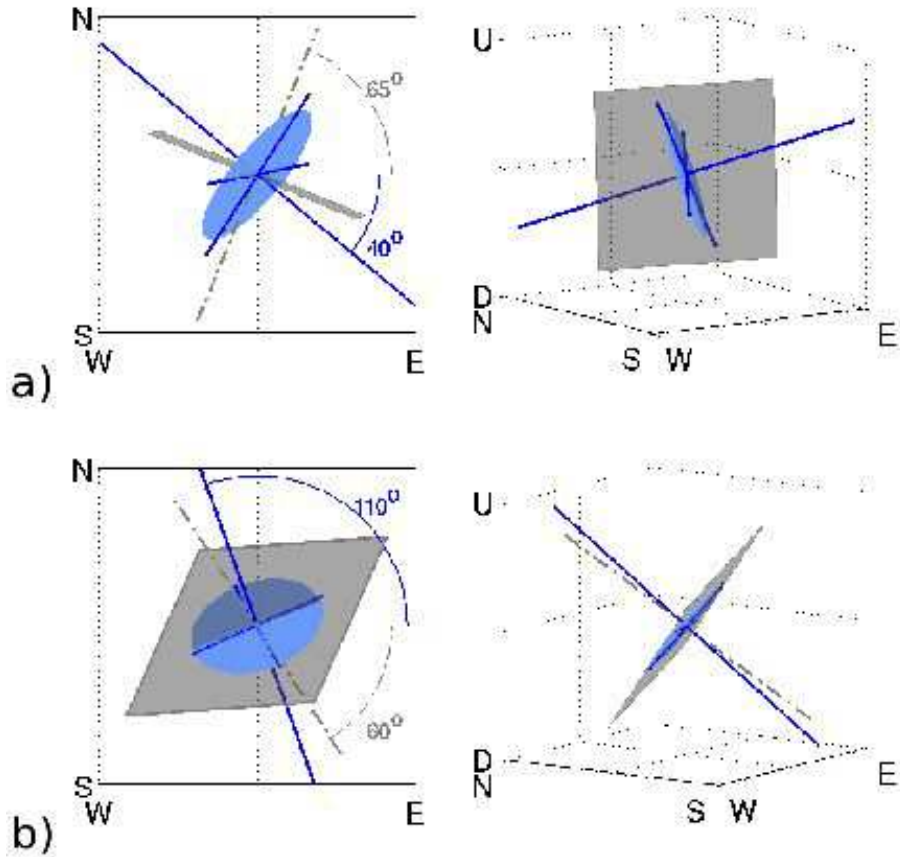}
\caption{Source mechanisms obtained for a) Family 1 and b) Family 2 events,  for the crack constrained inversion (Cr+F). The left panels are map views, and the right panels are 3D views.  The circular areas represent  the cracks. The solid lines represent the normalized eigenvectors, the longest lines are the crack normals. The light grey squares show the location structures obtained by \cite{debarros09}, i.e. a) a sub-vertical dike striking SWS-NEN for Family 1 and b) a 45\degre inclined plane striking  SW-NE containing the two pipe-like bodies of Family 2. The dashed lines are the normal vectors of these two structures. The azimuths of the normal vectors are given on the map view.}\label{fig9}
\end{figure*}

The orientations of the crack normals are ($\phi$=-40\degrenospace, $\theta$=70\degrenospace) and ($\phi$=110\degrenospace, $\theta$=50\degrenospace) (see Figure \ref{fig9}), which correspond to cracks striking in the SW-NE and SWS-NEN directions. Uncertainty of the inversion is $\pm$10\degrenospace. Cracks of both families are roughly orthogonal but their strikes show a similar orientation (between N40\degre E and N70\degre E). The orientation of these cracks are different from the crack found by \citet{lokmer07b} ($\phi$=35\degre and $\theta$=72\degrenospace). However, those events were recorded during the 2004 eruption and showed different waveforms and spectral characteristics.  As expected, the LP seismicity on Mt Etna is not constant, both for the seismicity rate and for the source properties \citep{patane08}.\\

For Family 1, the MT solution (crack) is roughly orthogonal to the source area determined by \citet{debarros09}, which is a subvertical dike-like structure striking NW-SE (see Fig. \ref{fig9}).  For Family 2, the crack solution lies on the plane (with normal oriented by $\phi$=120\degrenospace and $\theta$=45\degrenospace) containing the two pipe-like bodies. The LP events are not necessarily produced by the structure in which they are located. As expected, the inversion of all the individual events (see Fig. \ref{fig5} and \ref{fig6}) shows that the mechanism is highly repetitive.\\

% In this case, the main peak is negative, the mechanism is associated to a closure of the crack.
For both families, the source time function is very short (i.e. less than 4 s), which suggests a pulsing rather than an oscillating mechanism. The STFs can be seen as an impulsive function filtered in the frequency band of interest (0.2-1.2Hz). Amplitudes of the seismic moment are about 43 and 25 $10^9$ N.m for Family 1 and 2, respectively. Volumetric change $\Delta V$ can be estimated from $M_0=\mu~\Delta V$. From the velocity of the medium we compute a rough approximation of the dynamic shear modulus: $\mu=2.9$ GPa, which we assume to  be equivalent to the static one. The volume changes are  15 and 9 m$^3$ respectively. These volumes are smaller than the one found by \cite{lokmer07b}, but they are in agreement with the lower amplitude of the signals and the shallower source positions.  These volumes correspond to a normal displacement of 1 mm for a 100 meter sided square crack and 10 cm for a 10 m one.  \\

\subsection{Relationship to the eruption}
\cite{debarros09} and this study show that 1) two families of LP events are spread out along structures located between 800 m and the surface, 2) their source mechanisms are related to cracks, with orientations which are not necessarily the same as the structures along which the events are located, 3) for both families, crack strikes are roughly similar while dips are orthogonal, and  4) signals and source mechanisms are similar within each family.\\

Following the lava fountain of the 10$^{th}$ of May 2008, an eruptive fissure opened on the eastern flank of the volcano.  During the experiment time, lava flowed below the LP source locations at  a distance of 1.5 km. Because only gases were being released from the summit craters, it is possible that no magma had risen to the upper part of Mount Etna. This suggests that LP events are maybe not directly related to magma movements. Moreover,  gases may be the most likely fluids present in the main conduits and in the fractures surrounding them. \\

The lava fountain at the beginning of the eruption was associated with high fluid pressure that can destabilize the edifice by opening fractures in the upper part of the volcano. After this event, the flank lava flow and the summit degassing certainly drained the cone producing a decrease of the pressure. The LP events may be linked to this decompressive phase.  The decrease of pressure can lead to instabilities of those fractures as the volcano settled under its own weight. \cite{patane08} analysed a family of LP events (called Family 2 in \cite{patane08}) occurring only after the lava fountains of 2007, which have similar characteristics to the events studied here. They lasted for approximately one month after the lava fountain and were interpreted as the response to the volcano deflation. This is also confirmed by \cite{falsaperla02} who linked the LP activities to the collapses of the crater floor. In this case, the fluids involved can be gas or steam. Gases contained in the cracks are suddenly expelled to the main conduits. This can produce LP events with mechanisms similar to hydraulic transients \citep{Ferrick82} or hydrodynamic instabilities of nonlaminar flows \citep{Rust2008}.  This hypothesis can be linked to the laboratory studies performed by \cite{benson08}, who show that the decompression phase in rock samples can generate LP events in complex-shape fractures belonging to the damage zone of the main conduits. Similarly to our study, the events are located in the main conduits but are not directly generated inside them.\\ 

Cracks for both families are striking SW-NE. The orientation of the cracks and of the location structures are consistent with the tectonic setting, which generate faults in the NW-SE and SW-NE direction \citep{bonnacorso04}. However, as they are in the shallowest part of the volcano, they are more likely due to gravity effects. In particular, the East flank of the volcano is collapsing and successive eruptions strongly destabilized this area. This generates weaknesses oriented SW-NE, which are coherent with the azimuth of the cracks.  The cessation of the LP events after the 22$^{nd}$ of June suggests that the upper part of the volcano reached an equilibrium, where pressure and stress return to a static state. The decompression phase, following the lava fountain of the 10$^{th}$ of May, 2008, lasts about 40 days, which is in agreement with the conclusion of \cite{patane08}.    \\

\section{Conclusion}
Two families of LP events, comprising  63 and 66 events respectively, are selected from the first four days of a seismic experiment on Mount Etna (18/06/2008-03/07/2008). 50 stations, including 30 stations located in close proximity to the summit, were used for this study. Moment tensor inversion of numerical data shows that, for this deployment,  it is more reliable to use forces in the inversion to correctly describe the moment. However, the forces cannot be correctly reconstructed as they strongly reflect the errors coming from the velocity mismodelling and source mislocation. In general, as MTI appears strongly sensitive to station distribution, numerical tests are therefore required before every MTI study. Stations close to the source positions are required to correctly invert  Long Period events. \\

We perform moment tensor inversions in two steps. First, we determine the type of mechanism involved using unconstrained inversion. We then constrain the inversions to this particular mechanism to confirm the solution and refine its characteristics. Inversions of the events  of the two families show mechanisms with high volumetric components,  most likely generated by cracks. For both families, cracks are striking in the SW-NE direction, while their dips are approximately orthogonal to each other. The crack orientations are thus different from the location structures obtained by \citet{debarros09}. This can suggest that the LP events are generated by the fractures which belong to the damaged zone around the main conduits of the volcano. We hypothesize that these events are related to the decompression phase following the lava fountain of the 13$^{th}$ of May, and not to the lava flow from the flank eruption.\\

MTI reveals strong forces, especially for Family 2, but we are not able to determine if they are real or due to artefacts in the moment tensor inversion. These forces, as well as the time shifts observed between the moment tensors components, can be due  to uncertainties in the velocity model or to complex sources (e.g., dual sources, torque)  that cannot be accurately reconstructed by second-order MTI as used here. To solve this problem, a better knowledge of the velocity model is needed. To be able to unambiguously explain both moment and forces, a more general approach must take into account: i) extended sources, ii) multiple sources, and iii) rotational effects. \\

\begin{acknowledgments}
L. De Barros and G.S. O’Brien were part funded by the department of Communications, Energy and Natural Resources (Ireland) under the National Geosciences programme 2007 – 2013. Financial
assistance for field work from University College Dublin and the EU VOLUME project, and computational facilities from SFI/HEA ICHEC
are acknowledged. We wish to thank M. M\"ollhoff, A. Braiden, J. Grangeon, P. Bascou, M. La Rocca, D. Galluzzo and S. Rapisarda for assistance in the field experiment. The relevant comments made
by the associate editor and the reviewers Philip M. Benson, Jan \v{S}\`ilen\`y and Philippe Jousset greatly contributed to improve the manuscript. 
\end{acknowledgments}

\newpage
\begin{table}[p]
\caption{Misfit (Mis) in percentage and moment eigenvalues (MTE) for different constrained (Cr, Pi, Ex denote crack, pipe and explosion constrained inversion, respectively) or unconstrained (MT) inversion with and without single forces (F). The true mechanism involved vertical ($C_X$) and inclined ($C_{L}$) cracks, as well as vertical ($F_Z$) and inclined ($F_{45}$) single forces.}\label{tab1}
\vspace{0.5cm}
\begin{center}
\begin{tabular}{|c|c|c||c|c||c|c||c|c||c|c|}
\hline
\multicolumn{2}{|c|}{True Mech.} & & \multicolumn{2}{c||}{Unconstrained}& \multicolumn{2}{c||}{Crack} & \multicolumn{2}{c||}{Pipe}& \multicolumn{2}{c|}{Explosion}\\ 
\hline
MT & F &  & MT & MT+F & Cr & Cr+F & Pi &  Pi+F & Ex & Ex+F \\ 
\hline
\hline
 \multirow{2}{1cm}{\centering $C_X$} & \multirow{2}{1cm}{\centering /} & Mis & 29 & 24 & 46 & 28 & 52 & 47 & 69 & 63 \\ 
& & MTE & 1:1.2:2.4 & 1:1.1:2.8  & 1:1:3 & 1:1:3 & 1:2:2 & 1:2:2 & 1:1:1 & 1:1:1 \\ 
\hline
\hline
 \multirow{2}{1cm}{\centering $C_X$} & \multirow{2}{1cm}{\centering $F_{45}$} & Mis& 47  & 27 & 63 &37 & 56 & 39 & 74 & 54 \\ 
& & MTE & 1:1.8:2.2  & 1:1.4:3.4 & 1:1:3 & 1:1:3 & 1:2:2 & 1:2:2 & 1:1:1 & 1:1:1 \\ 
\hline
\hline
 \multirow{2}{1cm}{\centering $C_X$} & \multirow{2}{1cm}{\centering $F_{Z}$} & Mis& 30  & 21 & 61 &27 & 40 & 36 & 63 & 54 \\ 
& & MTE & 1:1.6:2.1  & 1:1.4:3.2 & 1:1:3 & 1:1:3 & 1:2:2 & 1:2:2 & 1:1:1 & 1:1:1 \\ 
\hline
\hline
 \multirow{2}{1cm}{\centering $C_L$} & \multirow{2}{1cm}{\centering /} & Mis& 33  & 18 & 54 &21 & 60& 49 & 75 & 68 \\ 
& & MTE & 1:1.3:2.0  & 1:1:2.5 & 1:1:3 & 1:1:3 & 1:2:2 & 1:2:2 & 1:1:1 & 1:1:1 \\ 
\hline
\hline
 \multirow{2}{1cm}{\centering $C_L$} & \multirow{2}{1cm}{\centering $F_{45}$} & Mis& 53  & 19 & 68 &22 & 63 & 25 & 70 & 31 \\ 
& & MTE & 1:1.4:1.8  & 1:1.4:3.4 & 1:1:3 & 1:1:3 & 1:2:2 & 1:2:2 & 1:1:1 & 1:1:1 \\ 
\hline
\hline
 \multirow{2}{1cm}{\centering $C_L$} & \multirow{2}{1cm}{\centering $F_{Z}$} & Mis& 39  & 20 & 75 &24 & 48 & 38 & 63 & 51 \\ 
& & MTE & 1:1.5:1.8  & 1:1.1:2.4 & 1:1:3 & 1:1:3 & 1:2:2 & 1:2:2 & 1:1:1 & 1:1:1 \\ 
\hline
\end{tabular}
\end{center}
\end{table}

\begin{table}[p]
\caption{Misfit value (Mis),  moment tensor eigenvalues (MTE) for both families and different inversion.} \label{tab2}
\vspace{0.5cm}
 \begin{center}
\begin{tabular}{|c|ccc|}
\hline
Family & Inv & Mis (\%) & MTE \\ 
\hline
 \multirow{3}{1cm}{\centering F1} & MT+F & 27  & 1:1.1:2.2 \\ 
 & MT &  61 & 1:1.5:3.2\\
& Cr+F &  38 & 1:1:3\\
& Cr &  78 & 1:1:3\\ 
\hline
\multirow{6}{1cm}{\centering F2}& MT+F & 21   & 1:1.1:1.6\\ 
  & MT &  67& 1:3.1:3.8 \\ 
 & E+F &  42  & 1:1:1\\ 
 & Pi+F &  40  & 1:2:2\\ 
& Cr+F &38  & 1:1:3\\ 
& Cr &  84  & 1:1:3\\ 
 \hline
\end{tabular}
\end{center}
\end{table}

\newpage

\end{article}

\end{document}